\documentclass[final,5p,times,sort&compress]{elsarticle}

\usepackage{amsmath}
\usepackage{mathrsfs}
\usepackage{bm}
\usepackage[colorlinks]{hyperref}

\journal{Physics Letters B}

\begin{document}

%%%%%%%%%%%%%%%%%%%%%%%%%%%%%%%%%%%%%%%%%%%%%%%%%%%%%%%%%%
% user definitions

\let\de=\partial
\let\eps=\epsilon
\let\eqt=\triangleq
\newcommand\La{\mathscr{L}}
\newcommand\imag{\mathrm i}
\newcommand\dd{\mathrm d}
\newcommand\vek[1]{\bm{#1}}
\newcommand\VAR[2]{\frac{\delta #1}{\delta #2}}

%%%%%%%%%%%%%%%%%%%%%%%%%%%%%%%%%%%%%%%%%%%%%%%%%%%%%%%%%%

\begin{frontmatter}

\title{Spontaneous breaking of spacetime symmetries and the inverse Higgs effect}

\author[helsinki,rez]{Tom\'a\v{s} Brauner\corref{cor}}
\ead{tomas.brauner@helsinki.fi}
\cortext[cor]{Corresponding author.}
\author[berkeley]{Haruki Watanabe}

\address[helsinki]{Department of Physics, University of Helsinki, 00560 Helsinki, Finland}
\address[rez]{Department of Theoretical Physics, Nuclear Physics Institute of the ASCR, 25068 \v{R}e\v{z}, Czech Republic}
\address[berkeley]{Department of Physics, University of California, Berkeley, California 94720, USA}

\begin{abstract}
It has been long known that when spacetime symmetry is spontaneously broken, some of the broken generators may not give rise to independent gapless, Nambu--Goldstone excitations. We provide two complementary viewpoints of this phenomenon. On the one hand, we show that the corresponding field degrees of freedom have the same symmetry transformation properties as massive, matter fields. The ``inverse Higgs constraints'', sometimes employed to eliminate these modes from the theory, are reinterpreted as giving a field parametrization that makes these transformation properties manifest. On the other hand, relations among classical symmetry transformations generally lead to identities for the associated Noether currents that allow saturation of the Ward--Takahashi identities for all the broken symmetries with fewer gapless excitations than suggested by mere counting of broken generators.
\end{abstract}


\end{frontmatter}

%%%%%%%%%%%%%%%%%%%%%%%%%%%%%%%%%%%%%%%%%%%%%%%%%%%%%%%%%%

\section{Introduction}

Spontaneous breakdown of a global continuous symmetry leads to gapless excitations in the spectrum: the Nambu--Gold\-stone (NG) bosons. A detailed analysis of broken \emph{internal} symmetries in relativistic field theory now belongs to standard textbook material. On the contrary, comprehensive understanding of NG bosons of internal symmetries in the nonrelativistic context has been reached only very recently~\cite{Watanabe:2011ec,Watanabe:2012hr,Hidaka:2012ym}. In this paper, we deal with some questions pertinent to \emph{spacetime} symmetries.

An invaluable tool for building invariant Lagrangians for NG bosons is provided by the coset construction~\cite{Coleman:1969sm,Callan:1969sn}. Together with the proof that in Lorentz-invariant theories, the effective Lagrangian density can always be made invariant by a suitable choice of field variables~\cite{Leutwyler:1993iq}, this establishes a complete, exhaustive framework for construction of effective field theories (EFTs) for internal symmetries. The nonrelativistic case is subtle though, as the effective Lagrangian is not necessarily invariant~\cite{Leutwyler:1993gf}. Even though one then cannot apply the coset construction quite straightforwardly, it is still possible to express the most general effective Lagrangian in terms of its basic building block, the Maurer--Cartan (MC) form~\cite{Watanabe:2013uya}.

The coset construction was soon extended to spacetime symmetries~\cite{Volkov:1973vd,Ogievetsky}. This generalization was designed for, and with rare exceptions applied to, situations where translational invariance is not spontaneously broken. Yet, there seems to be no proof analogous to Ref.~\cite{Leutwyler:1993iq} that, even with this restriction, the coset construction actually yields the most general EFT, consistent with all the symmetries. Moreover, there are indications that for broken translations, it requires modification.

It is well known~\cite{Volkov:1973vd,Ivanov:1975zq} that when several spacetime symmetries are spontaneously broken at the same time, some of them may not give rise to independent NG modes. A common example is a crystalline solid where both spatial translations and rotations are broken, yet only the former produce NG bosons, the acoustic phonons. Within the coset construction, where one field variable is introduced for each broken symmetry generator, this is taken into account by an operational prescription known as the inverse Higgs mechanism~\cite{Ivanov:1975zq}. This allows one to remove the ``unwanted'' degrees of freedom by imposing a set of constraints, compatible with all the symmetries.

The physical interpretation of the inverse Higgs constraints (IHCs) seems to be ambiguous in the literature. As emphasized recently in Ref.~\cite{Nicolis:2013sga}, having a mere \emph{possibility} to eliminate some degrees of freedom is not satisfactory. One should not have the freedom to impose such constraints at will, even if in some cases doing so may turn out equivalent to using the equations of motion~\cite{McArthur:2010zm}. Our discussion in Sec.~\ref{sec:higgs} elaborates the arguments of Ref.~\cite{Nicolis:2013sga} and hopefully clarifies this issue.

The absence of independent NG modes for some spacetime symmetries can be traced back to the fact that local forms of certain spacetime transformations, such as translations and rotations, coincide~\cite{Low:2001bw}. This intuitive, but essentially classical, argument has recently been generalized and reformulated in a quantum-field-theoretic language~\cite{Watanabe:2013iia}, see also recent Ref.~\cite{Hayata:2013vfa}. In Sec.~\ref{sec:noether}, we further detail this viewpoint. We observe that the Ward--Takahashi identities for different broken symmetries can be saturated by the same NG boson as long as the associated Noether currents are related, thus providing an alternative insight in the nature of NG bosons of spacetime symmetries.

%%%%%%%%%%%%%%%%%%%%%%%%%%%%%%%%%%%%%%%%%%%%%%%%%%%%%%%%%%

\section{Summary of the results}
\label{sec:summary}

The question of finding the most general low-energy effective Lagrangian for the NG bosons is equivalent to classifying all nonlinear realizations of the broken symmetry~\cite{Coleman:1969sm,Callan:1969sn}. We do not attempt to give a formal proof that the coset construction actually gives the most general EFT in case of broken spacetime symmetries. However, we show that:
\begin{enumerate}[(i)]
\item Within the coset construction, there is a field parametrization in which a given nonlinear realization takes the form of certain minimal realization augmented by additional non-NG, ``matter" degrees of freedom.
\item The minimal nonlinear realization is given solely by the symmetry-breaking pattern. Hence the number of true, gapless NG bosons is independent of the microscopic me\-cha\-nism that causes the symmetry breaking.
\item Relations among classical local symmetry transformations lead generally to identities among the associated Noether currents. These establish a direct link between the symmetry group and the spectrum of NG modes.
\end{enumerate}
Our first two points show that the question whether the ``redundant'' NG modes should, or should not, be eliminated from the EFT by imposing the IHCs is to some extent moot. The ``constraints'' provide us with a field parametrization in which these modes transform as matter fields. This leads to an unambiguous conclusion that they are not true, massless NG bosons, but rather massive, non-NG ones. Whether they actually appear in the spectrum of the theory is a dynamical question which cannot be answered within the EFT alone.

%%%%%%%%%%%%%%%%%%%%%%%%%%%%%%%%%%%%%%%%%%%%%%%%%%%%%%%%%%

\section{Toy example}
\label{sec:toy}

Consider the theory of a free massless relativistic scalar field $\phi$, defined by $\La=\frac12(\de_\mu\phi)^2$. Its action is invariant, apart from the spacetime symmetries, under a shift of the field, $\phi'(x)=\phi(x)+\alpha+\beta_\mu x^\mu$. In the ground state, both the $\alpha$- and the $\beta$-shift are spontaneously broken, giving altogether $2+d$ broken generators in $d$ spatial dimensions. However, there is obviously only one NG mode in the spectrum. There are different ways to understand why the $\beta$-shifts do not give rise to additional $1+d$ NG modes. The NG excitations can be viewed as spacetime-dependent fluctuations of the order parameter, generated by local broken symmetry transformations. Now the local forms of the $\alpha$- and the $\beta$-shift coincide, or more precisely, the latter can be reproduced by the former by setting $\alpha(x)=\beta_\mu(x)x^\mu$. Hence the fluctuations induced by the $\beta$-shifts are not independent, in line with the argument of Ref.~\cite{Low:2001bw}.

Within the coset construction, one introduces one NG field for each broken generator, here $\pi$ for the $\alpha$-shift and $\rho_\mu$ for the $\beta$-shifts. Following the general setup applicable to spacetime symmetries~\cite{Volkov:1973vd,Ogievetsky}, the coset element has to take into account all nonlinearly realized symmetries including spacetime translations, and can be written as $U(x)=e^{\imag x^\mu P_\mu}e^{\imag\pi(x)M}e^{\imag\rho_\mu(x)N^\mu}$, where $P^\mu$ is the energy--momentum operator whereas $M$ and $N^\mu$ are the generators of the field shifts. The only nontrivial commutator among these generators is $[P_\mu,N_\nu]=\imag\eta_{\mu\nu}M$ where $\eta_{\mu\nu}$ is the Minkowski metric. From here, one deduces the transformation of the NG fields under the action of the internal symmetry group, defined by left multiplication of $U$ by $e^{\imag\alpha M}e^{\imag\beta_\mu N^\mu}$,
\begin{equation}
\pi'(x)=\pi(x)+\alpha+\beta_\mu x^\mu,\qquad
\rho'_\mu(x)=\rho_\mu(x)+\beta_\mu.
\end{equation}
The MC form is defined as usual as $\omega=-\imag U^{-1}\dd U$ and reads
\begin{equation}
\omega=\dd x^\mu P_\mu+(\dd\pi-\rho_\mu\dd x^\mu)M+\dd\rho_\mu N^\mu.
\label{omegatoy}
\end{equation}
In the standard inverse Higgs philosophy~\cite{Ivanov:1975zq}, we could now set the $M$-component of the MC form to zero. This is a condition invariant under all the symmetries of the problem, allowing us to eliminate $\rho_\mu$ in favor of $\pi$, namely $\rho_\mu=\de_\mu\pi$. We are left with the invariant vielbein, $\omega_P^{\mu\nu}=\eta^{\mu\nu},$ and the NG part of the MC form, $\omega_N^{\mu\nu}=\de^\mu\rho^\nu=\de^\mu\de^\nu\pi$. Note that $\omega^{\mu\nu}_N$ cannot be used to construct the original Lagrangian density $\La$ in the usual way. The reason for this is that $\La$ changes by a surface term upon a $\beta$-shift. The Lagrangian can, however, be obtained in an analogy with the Wess--Zumino--Witten term~\cite{Goon:2012dy}.

We next explain how the elimination of the ``redundant'' degree of freedom $\rho_\mu$ can be circumvented. Eq.~\eqref{omegatoy} hints that upon the field redefinition $\rho_\mu\to\Omega_\mu=\rho_\mu-\de_\mu\pi$, the new field $\Omega_\mu$ will transform covariantly (in this model it is even invariant). Consequently, it is possible to add a mass term $\Omega_\mu^2$ to the effective Lagrangian, consistent with all the symmetries. The field $\Omega_\mu$ does not represent a NG boson! This is as much as we can get from the symmetry-breaking pattern alone. How the effective Lagrangian including both $\pi$ and $\Omega_\mu$ looks like depends on how we \emph{define} power counting in our EFT. Depending on the size of the mass $M_\Omega$ of $\Omega_\mu$, two natural schemes are conceivable:
\begin{enumerate}[I.]
\item $M_\Omega$ is order-$0$ in the derivative expansion. At the leading order, $\Omega_\mu$ only has a mass term but no kinetic term, hence it does not describe a propagating mode. In this case, the leading-order equation of motion can be equivalent to imposing the IHC, $\Omega_\mu=0$, as observed in Ref.~\cite{McArthur:2010zm}.
\item $M_\Omega$ is order-$1$ in the derivative expansion. Here the kinetic and mass terms of $\Omega_\mu$ are of the same order and hence $\Omega_\mu$ describes a propagating massive mode in the EFT. At which order of the derivative expansion it enters depends on the assumption one makes for the scaling of $\Omega_\mu$ itself.
\end{enumerate}
In any case, the EFT itself cannot tell us whether $\Omega_\mu$ excites an independent physical state. In scheme I, $\Omega_\mu$ is simply an auxiliary field which couples to the same state as $\pi$. However, even in scheme II, the mass $M_\Omega$ can, if somewhat artificially, be taken to infinity, thus eliminating $\Omega_\mu$ from the observable spectrum. While in the present trivial model, there is obviously no massive particle described by $\Omega_\mu$, in the next section, we will suggest some examples where $\Omega_\mu$ may be physical though.

This example allows us to draw several morals, which apply also to the general case discussed in the next section. First, \emph{eliminating the massive field by imposing the IHC and by using the equation of motion may result in apparently different EFTs}. This is obviously the case for our scheme II. One should nevertheless emphasize that how precisely the redundant degree of freedom is removed from the theory is immaterial; see the discussion in Sec.~\ref{subsec:discussion} for more details.

Second, within scheme II, \emph{imposing the IHC modifies the physical content of the EFT}. We start with an EFT containing a NG boson $\pi$ and a massive mode $\Omega_\mu$, and upon imposing the IHC end up with a theory of $\pi$ alone. Nevertheless, both theories are equally good EFTs for the massless relativistic scalar that we started with: at energy scales below $M_\Omega$, $\Omega_\mu$ will decouple anyway. 

Third, \emph{these arguments do not rely on the use of the coset construction to build the effective Lagrangian}. We only needed the MC form~\eqref{omegatoy} to identify the field $\Omega_\mu$ with suitable transformation properties. Apart from that, we can construct the effective Lagrangian for instance by collecting all possible operators allowed by the linearly realized symmetries, and then imposing the remaining symmetries by hand.

Finally, our simple toy model shows that \emph{the subtleties concerning the spectrum of NG modes are not limited to spacetime symmetries}, but apply more generally to \emph{nonuniform} symmetries, that is roughly, symmetries that do not commute with spacetime translations (see Ref.~\cite{Watanabe:2011ec} for a precise definition). 

%%%%%%%%%%%%%%%%%%%%%%%%%%%%%%%%%%%%%%%%%%%%%%%%%%%%%%%%%%

\section{Inverse Higgs constraints}
\label{sec:higgs}

We are now ready to generalize what we learned by studying the simple toy model. There are two ways to construct the low-energy EFT for a given system. The first one starts from a known microscopic theory and eliminates all degrees of freedom but the NG modes. The second approach requires to construct the most general effective action for the NG degrees of freedom alone; its a priori undetermined effective couplings are to be fixed either by experiment or by matching to the microscopic theory. We will discuss these approaches separately.

%%%%%%%%%%%%%%%%%%%%%%%%%%%%%%%%%%%%%%%%%%%%%%%%%%%%%%%%%%

\subsection{Microscopic picture}
\label{subsec:microscopic}

A key step in integrating out the massive modes from the microscopic theory is finding a suitable field parametrization that makes the NG degrees of freedom manifest. Let us recall how one proceeds in the case of uniform internal symmetries~\cite{Weinberg:1996v2}. Suppose that the order parameter for breaking the symmetry group $G$ down to its subgroup $H$ is given by an expectation value of an elementary (not necessarily scalar) field, $\langle\phi(x)\rangle=\phi_0$. (The field can carry indices, both internal and spacetime, which are not displayed.) A general point in the field space can then be expressed as
\begin{equation}
\phi(x)=U(x)[\phi_0+\chi(x)],
\label{microscopic}
\end{equation}
where $U(x)$ is a local $G$-transformation and $\chi(x)$ is a field from which all NG degrees of freedom have been eliminated. This can be ensured by imposing the condition $\mathrm{Im}(\chi^\dagger T_i\phi_0)=0$, where $T_i$ are the generators of $G$.

The matrix $U(x)$ is only defined up to a right multiplication by an element of $H$, and it is common to parametrize it as $U(x)=e^{\imag\pi^a(x)T_a}$ in terms of NG fields $\pi^a(x)$ and the broken generators of $G$, $T_a$. This way, one can obtain a \emph{one-to-one} field parametrization for any uniform internal symmetry. When the massive modes are simply dropped, $\phi(x)=e^{\imag\pi^a(x)T_a}\phi_0$, one arrives  at a particular EFT known as the nonlinear sigma model.

For nonuniform symmetries, on the contrary, this prescription does not lead to a one-to-one field parametrization. In the example from the previous section, the field $\phi(x)$ can obviously be obtained by a local shift generated by $M$, acting on the vacuum, $\phi_0=0$. Formally, this can be written as $\phi(x)=e^{\imag\pi(x)M}\phi_0$. As observed before, local fluctuations $\rho_\mu(x)$ induced by $N^\mu$ are not independent; they can be completely removed by a redefinition of $\pi(x)$. As another example, consider a two-dimensional crystal lattice. The spontaneously broken generators in this case include translations $P_x,P_y$ and the rotation $J$. Using the well-known relation $J=xP_y-yP_x$, the would-be rotation NG mode $\theta(x)$ can be eliminated in favor of the translation NG modes, as follows from $e^{\imag\theta J}=e^{\imag[(x\theta)P_y-(y\theta)P_x]}$. The origin and importance of such relations among symmetry generators will be discussed at length in Sec.~\ref{sec:noether}.

In Ref.~\cite{Nicolis:2013sga}, this redundancy is interpreted as a sort of gauge freedom: one starts by introducing one NG field for each spontaneously broken generator and then notices that the corresponding fluctuations of the order parameter are not independent. The parametrization of the microscopic field $\phi(x)$ can be made one-to-one by ``fixing the gauge'', that is, by setting some of the NG fields to zero. From the point of view of the microscopic theory, however, it is obviously more natural to work with a one-to-one field parametrization to start with.

%%%%%%%%%%%%%%%%%%%%%%%%%%%%%%%%%%%%%%%%%%%%%%%%%%%%%%%%%%

\subsection{Effective field theory picture}
\label{subsec:EFT}

Within the coset construction, one builds a model-in\-de\-pen\-dent low-energy EFT using the information about the symmetry breaking alone. Formally, this means that the dynamical degrees of freedom are given by the matrix $U(x)$ itself rather than its action on a particular order parameter $\phi_0$. Hence, all the fields $\pi^a(x)$ \emph{are} independent in this picture. For nonuniform symmetries, the translations have to be included in the definition of the coset element, and we thus set $U(x)=e^{\imag x^\mu P_\mu}e^{\imag\pi^a(x)T_a}$. The action of a transformation $g$ from the symmetry group $G$, now including spacetime transformations, is defined by~\cite{Ogievetsky}
\begin{equation}
ge^{\imag x^\mu P_\mu}e^{\imag\pi^a(x)T_a}=e^{\imag x'^\mu P_\mu}e^{\imag\pi'^a(x')T_a}h(g,\pi(x)),
\label{gtransfo}
\end{equation}
where $h\in H$ is a compensating transformation which ensures that $U$ remains in the coset space $G/H$. Denoting the unbroken generators of $H$ (apart from the momentum operator $P_\mu$) as $T_\rho$, the MC form, defined by $\omega=-\imag U^{-1}\dd U$, can be expanded as
\begin{equation}
\omega=\omega_P+\omega_\perp^aT_a+\omega_\parallel^\rho T_\rho,
\end{equation}
where $\omega_P=e_\nu^{\phantom\nu\mu}P_\mu\dd x^\nu$. The individual components of the MC form then transform in accord with Eq.~\eqref{gtransfo} as
\begin{equation}
\omega_{P,\perp}\to h\omega_{P,\perp}h^{-1},\qquad
\omega_\parallel\to h\omega_\parallel h^{-1}-\imag h\dd h^{-1}.
\end{equation}
If necessary, the EFT can be augmented with other, non-NG degrees of freedom, usually called \emph{matter} fields. Consider a field $\psi$ that transforms in some linear representation $R$ of the unbroken subgroup $H$, $\psi'(x')=R(h)\psi(x)$. This can be extended using Eq.~\eqref{gtransfo} to a nonlinear realization of the whole group $G$,
\begin{equation}
\psi'(x')=R(h(g,\pi(x)))\psi(x).
\label{matter}
\end{equation}
An invariant action can be constructed using as building blocks the covariant derivatives of NG fields, extracted from $\omega_{\perp}^a=e_\nu^{\phantom\nu\mu}D_\mu\pi^a\dd x^\nu$, the matter fields $\psi$, their covariant derivatives, defined by $\mathrm D\psi=(\dd+\imag\omega_\parallel)\psi=\omega_P^\mu D_\mu\psi$, and the volume measure, obtained from the invariant vielbein $e_\nu^{\phantom\nu\mu}$~\cite{Ogievetsky}. Using the definition of the coset element $U(x)$, the broken part of the MC form $\omega_\perp$ can be expressed, to linear order in the NG fields, as
\begin{equation}
\omega_{\perp}^a=(\de_\mu\pi^a-f^a_{\mu b}\pi^b)\dd x^\mu+\dotsb,
\end{equation}
where the structure constant is defined by $[P_\mu,T_b]=\imag f^a_{\mu b}T_a$.

Suppose now that there are two sets of broken generators, $T_\alpha$ and $T_A$, that transform in some representations $R_1$ and $R_2$ of $H$, $R_2$ being irreducible. Let us further assume that the translation generators $P_\mu$ span a representation $R_P$ of $H$, and that a direct product of any two representations of $H$ is fully reducible. Provided that the direct product $R_P\otimes R_1$ contains $R_2$ in its decomposition, and that some of the structure constants $f^\alpha_{\mu A}$ is nonzero, the projection of the covariant derivative $D_\mu\pi^\alpha$ to $R_2$, accomplished with the appropriate Clebsch--Gordan coefficients and denoted as $D_\mu\pi^\alpha\rvert^A$, contains a term linear in $\pi^A$~\cite{Ivanov:1975zq}. In the standard inverse Higgs language, this means that the field $\pi^A$ can be eliminated from the theory by imposing the IHC
\begin{equation}
D_\mu\pi^\alpha\rvert^A=0.
\label{IHC}
\end{equation}
Here we propose a different interpretation. We instead perform a \emph{field redefinition} $\pi^A\to\Omega^A=D_\mu\pi^\alpha\rvert^A$. The possibility to eliminate all $\pi^A$s in $R_2$ via a set of IHCs, stemming from the existence of terms linear in $\pi^A$ in the covariant derivatives, automatically guarantees that this field redefinition is nonsingular. Moreover, $\Omega^A$ transforms under $G$ in the same way as $\omega_\perp$. It therefore behaves as a matter field in the adjoint representation of $H$. As such, it is not protected by symmetry from attaining a mass term, and hence does not represent a NG boson.

%%%%%%%%%%%%%%%%%%%%%%%%%%%%%%%%%%%%%%%%%%%%%%%%%%%%%%%%%%

\subsection{Discussion}
\label{subsec:discussion}

Here we wish to make several comments that elucidate the physical meaning of the above observation. Our argument generalizes the discussion of Sec.~\ref{sec:toy}.

(i) There is only a limited number of true NG modes that are protected by symmetry. They are exactly those of the $\pi^a$ fields which cannot be eliminated by any IHC. This implies one of our main conclusions that nonlinear realizations of the broken symmetry can be reduced to certain minimal realization, to which matter fields can be added at will if necessary. The minimal realization, and hence the number of true NG bosons, depends solely on the symmetry-breaking pattern.

(ii) An EFT for the true NG bosons alone can be obtained from one for all the $G/H$ coset fields in different ways: by setting the $\Omega^A$s to zero (that is by imposing all possible IHCs), by eliminating the $\Omega^A$s via their equations of motion, or by integrating them out from the functional integral. The fact that the $\Omega^A$s have a mass term guarantees that in any case, we obtain a local EFT with a well-defined derivative expansion. While these three procedures can lead to seemingly different EFTs, upon a proper matching of the effective couplings, they have to give the same prediction for any physical observable.

(iii) The fact that the $\Omega^A$ fields acquire a mass term does not necessarily mean they are unphysical. One generic possibility how they could represent a massive state in the physical spectrum might be to focus on the vicinity of a second-order phase transition between phases where spacetime symmetries are spontaneously broken according to different patterns. However, whether such a phase transition can be realized, or more generally whether the anticipated massive states actually exist, has to be investigated case by case.

%%%%%%%%%%%%%%%%%%%%%%%%%%%%%%%%%%%%%%%%%%%%%%%%%%%%%%%%%%

\subsection{Examples}

For an example of how the IHC works, consider a nonrelativistic superfluid (see Ref.~\cite{Endlich:2013spa} for a recent discussion of this system from the IHC perspective). The ground state breaks the U(1) symmetry associated with particle number $Q$ as well as the Galilei boosts $B^i$. We choose the  coset parametrization $U(x)=e^{\imag x^\mu P_\mu}e^{-\imag\vek v(x)\cdot\vek B}e^{-\imag\theta(x)Q}$. Using the commutators of boost generators with the Hamiltonian and with the charge $Q$, $[H,\vek B]=\imag\vek P$ and $[P^i,B^j]=\imag\delta^{ij}mQ$, we obtain
\begin{equation}
\omega=\dd x^\mu P_\mu+\vek v\dd t\cdot\vek P-\dd\vek v\cdot\vek B-\Bigl[(\vek\nabla\theta+m\vek v)\cdot\dd\vek x+\bigl(\de_0\theta-\tfrac12{m\vek v^2}\bigr)\dd t\Bigr]Q.
\end{equation}
Assuming only $s$-wave interactions, which is appropriate at low particle densities, the superfluid can be described by a microscopic model for a complex Schr\"odinger field $\psi(x)$,
\begin{equation}
\La=\frac{\imag}{2}(\psi^\dagger\de_0\psi-\de_0\psi^\dagger\psi)-\frac{\vek\nabla\psi^\dagger\cdot\vek\nabla\psi}{2m}-\frac{g}{2}(\psi^\dagger\psi-n_0)^2.
\label{superfluid}
\end{equation}
By parametrizing the field as $\psi(x)=\sqrt{n(x)}\,e^{-\imag\theta(x)}$, one immediately notices that $n(x)$ is canonically conjugate to $\theta(x)$, so that the Lagrangian~\eqref{superfluid} describes just one degree of freedom. (This observation also excludes the existence of a Higgs mode in the nonrelativistic superfluid~\cite{Altman}.) Therefore, there is no space for the boost mode $\vek v(x)$ and it has to be eliminated from the EFT, for instance by using the IHC, $\vek\nabla\theta+m\vek v=\vek0$. One thereby obtains the effective Lagrangian,
\begin{equation}
\La=\mathcal{P}(\mu),\qquad\mu=\de_0\theta-\frac{m\vek v^2}{2}=\de_0\theta-\frac{(\vek\nabla\theta)^2}{2m},
\end{equation}
invariant under the Galilei transformation $\theta'(\vek{x}+\vek{v}t,t)=\theta(\vek{x},t)-m\vek{v}\cdot\vek{x}-\frac{1}{2}m\vek v^2t$. The unknown function $\mathcal P(\mu)$ can be shown to coincide with the pressure in thermodynamic equilibrium~\cite{Son}.

As the next example, we now discuss a superfluid in two spatial dimensions, rotating with angular velocity $\Omega$. In such a system, a triangular lattice of vortices spontaneously forms. Neglecting the centrifugal potential due to the rotation as well the trapping potential, the Lagrangian becomes identical to one for a system of charged bosons in a uniform magnetic field of intensity $eB=2m\Omega$. Let $\vek{X}$ be the body-fixed coordinates of the medium and $\vek{\Pi}$ the generator of the internal translation $\vek{X}'(\vek{x},t)=\vek{X}(\vek{x},t)+\vek{a}$~\cite{Son:supersolid}. The vortex lattice configuration breaks spontaneously both the spatial and the internal translations. Neglecting for simplicity the discreteness of the lattice and treating the system as an isotropic crystal, there is a residual continuous symmetry, generated by the diagonal translation operator $\vek P+\vek\Pi$. We now parametrize the coset as
\begin{equation}
U(x)=e^{\imag x^\mu P_\mu}e^{-\imag\vek X(x)\cdot\vek\Pi}e^{-\imag\theta(x)Q}.
\end{equation}
Here $\theta(x)$ represents the smooth part of the superfluid phase.  The vibrations of the vortex lattice are described by the displacement field, $\vek{u}(\vek{x},t)=\vek{x}-\vek{X}(\vek{x},t)$.

The nontrivial properties of collective modes in this system stem from the fact that the two components of momentum no longer commute with each other. The magnetic translation operator $\vek P$ satisfies $[P^i,P^j]=-2\imag m\Omega\eps^{ij}Q$ due to the effective magnetic field. Moreover, vortices make the internal translation $\vek\Pi$ noncommuting as well, $[\Pi^i,\Pi^j]=-2\imag\pi m_0\eps^{ij}Q$, where $m_0=-m\Omega/\pi$ is the number density of the vortices. Using these commutation relations, one obtains the MC form,
\begin{equation}
\omega=\dd x^\mu P_\mu-\dd\vek X\cdot\vek\Pi-\bigl[\dd\theta+m\Omega\eps_{ij}(x^i\dd x^j-X^i\dd X^j)\bigr]Q.
\end{equation}
The third term can be rewritten as
\begin{equation}
\mathcal{D}\theta=\dd\tilde\theta+m\Omega\eps_{ij}(2u^i\dd x^j-u^i\dd u^j),
\end{equation}
where $\tilde\theta=\theta+m\Omega\eps_{ij}x^iu^j$.  In this case, $\vek{u}(x)$ describes a \emph{physical gapped mode} with the gap $2m\Omega$~\cite{Watanabe:2013iia}, whose presence is guaranteed by Kohn's theorem~\cite{Kohn}. Therefore, imposing naively the IHC $\mathcal{D}_i\theta=0$, that is, $u^i=-\frac{1}{2m\Omega}\eps^{ij}\de_j\tilde\theta+\dotsb$, would clearly modify the physical content of the system. If one is only interested in the physics below this gap, one can integrate the displacement field out and get the effective Lagrangian for the phase fluctuation, $\varphi(\vek{x},t)=\mu_0t-\tilde\theta(\vek{x},t)$. The Lagrangian starts with $(\de_0\varphi)^2$ and $(\vek\nabla^2\varphi)^2$, so that this mode can be classified as type-II and type-A~\cite{Watanabe:2012hr,Horava}. In contrast to the usual type-II and type-B NG modes, such soft excitations tend to restore the broken symmetry even at zero temperature and the phase correlation shows only a quasi-long-range order (power law decay)~\cite{Sinova}.

%%%%%%%%%%%%%%%%%%%%%%%%%%%%%%%%%%%%%%%%%%%%%%%%%%%%%%%%%%

\section{Relations among Noether currents}
\label{sec:noether}

As remarked in Sec.~\ref{subsec:microscopic}, the redundancy in order parameter fluctuations can be explained as a consequence of certain relations among the broken generators. In this section, we will show that these follow directly from the properties of classical local symmetry transformations, our main result being Eq.~\eqref{constraint}.

%%%%%%%%%%%%%%%%%%%%%%%%%%%%%%%%%%%%%%%%%%%%%%%%%%%%%%%%%%

\subsection{Definition of a Noether current}

Consider a theory of a set of (not necessarily scalar) fields, denoted collectively as $\phi(x)$, defined by the action $S[\phi(x)]$. This is assumed to be a spacetime integral of a local Lagrangian density, which is a function of $\phi(x)$ and its derivatives. Suppose that the action is invariant under a simultaneous transformation of the fields and the spacetime coordinates,
\begin{equation}
\phi'(x')=\phi(x)+\eps\zeta(\phi(x),x),\qquad
x'=x+\eps\xi(x).
\label{symtransfo}
\end{equation}
Here $\eps$ is an infinitesimal parameter of the transformation. Let us now perform an infinitesimal transformation with a \emph{coordi\-na\-te-dependent} parameter $\eps(x)$ such that it reduces to Eq.~\eqref{symtransfo} for constant $\eps(x)=\eps$. Owing to the invariance assumption, the variation of the action has to take the form
\begin{equation}
\delta S=\int\dd x\,J^\mu(x)\de_\mu\eps(x),
\label{deltaS}
\end{equation}
which defines the associated Noether current $J^\mu(x)$. There are two sources of ambiguity in this definition. First, the extension of the symmetry transformation to local $\eps(x)$ is not unique. One could naively take the same functional form~\eqref{symtransfo} and merely replace $\eps\to\eps(x)$, but this is not mandatory. Any choice compatible with Eq.~\eqref{symtransfo} gives a current that is conserved for fields satisfying the equations of motion (\emph{on-shell}). Second, shifting the current by a vector function $G^\mu(x)$ such that $\de_\mu G^\mu=0$ for all field configurations (\emph{off-shell}) does not affect Eq.~\eqref{deltaS}. This kind of ambiguity gives rise to an equivalence relation that will be denoted by the symbol $\eqt$, that is, $J^\mu\eqt J^\mu+G^\mu$. 

%%%%%%%%%%%%%%%%%%%%%%%%%%%%%%%%%%%%%%%%%%%%%%%%%%%%%%%%%%

\subsection{Locally identical field transformations}

Consider two classes of symmetry transformations, characterized by infinitesimal parameters $\eps^\alpha_1$ and $\eps^a_2$. Assume that their (suitably chosen) local forms coincide, that is, there is a set of functions $f^a_\alpha(x)$ such that setting $\eps^a_2(x)=f_\alpha^a(x)\eps^\alpha_1(x)$ makes the two local transformations \emph{identical}. The variation of the action under such transformation reads
\begin{equation}
\delta S=\int\dd x\,J^\mu_{2a}\de_\mu\eps_2^a=\int\dd x\,J^\mu_{2a}(f_\alpha^a\de_\mu\eps^\alpha_1+\eps^\alpha_1\de_\mu f_\alpha^a).
\end{equation}
However, since $\eps^\alpha_1(x)$ is a parameter of an identical local transformation, this should be equal to $\delta S=\int\dd x\,J^\mu_{1\alpha}\de_\mu\eps_1^\alpha$. That is only possible provided
\begin{equation}
J^\mu_{2a}\de_\mu f_\alpha^a=\de_\mu N^\mu_\alpha
\label{Ncond}
\end{equation}
for some vector function $N^\mu_\alpha(x)$. By comparison, we then obtain
\begin{equation}
J^\mu_{1\alpha}\eqt f_\alpha^aJ^\mu_{2a}-N^\mu_\alpha.
\label{constraint}
\end{equation}
This alone would be a void statement (any two currents differ by some vector function) if it were not for the simultaneous condition~\eqref{Ncond} which must be satisfied off-shell. Thanks to this condition, the (on-shell) conservation of $J^\mu_{1\alpha}$ is a consequence of the conservation of $J^\mu_{2a}$. In fact, a stronger conclusion holds: $\de_\mu J^\mu_{1\alpha}=f^a_\alpha\de_\mu J^\mu_{2a}$ off-shell. Note that the argument leading to Eq.~\eqref{constraint} was previously used to derive the conserved currents associated with nonrelativistic Galilei invariance~\cite{Nicolis:2010se}.

Given the freedom that one has in the definition of the Noe\-ther currents, one may wonder if they can be ``improved'' so that the correction term $N^\mu_\alpha$ vanishes. This is indeed often possible, but cannot be achieved in general. Recall the free massless scalar field discussed in Sec.~\ref{sec:toy}. The Noether currents of the $\alpha$- and $\beta$-shifts are easily evaluated explicitly,
\begin{equation}
J^\mu\eqt\de^\mu\phi,\qquad
J^\mu_\alpha\eqt x_\alpha\de^\mu\phi-\eta^\mu_\alpha\phi.
\label{toycurrents}
\end{equation}
The local $\beta$-shift can be reproduced by a local $\alpha$-shift with $\alpha(x)=x_\alpha\beta^\alpha(x)$, that is, $f_\alpha(x)=x_\alpha$. We thus have $J^\mu\de_\mu f_\alpha=\de_\alpha\phi$, which implies $N^\mu_\alpha=\eta^\mu_\alpha\phi$; Eq.~\eqref{constraint} recovers the current of the $\beta$-shifts given in Eq.~\eqref{toycurrents}. However, it is not possible to improve the currents so that $\tilde J^\mu_\alpha\eqt x_\alpha\tilde J^\mu$ holds, since current conservation would imply $\tilde J_\alpha=0$ on-shell, in contradiction with the fact that the integral of the temporal component of the current is a generator of the associated symmetry transformation.

There are other, more important examples. Spacetime rotations acting on a scalar field correspond to setting $\zeta=0$ and $\eps\xi^\mu\to\theta^{\alpha\beta}\omega^\mu_{\alpha\beta}$ in Eq.~\eqref{symtransfo}, where $\theta^{\alpha\beta}$ is the antisymmetric matrix of parameters and $\omega^\mu_{\alpha\beta}=x_\alpha\eta^\mu_\beta-x_\beta\eta^\mu_\alpha$. A local rotation is then reproduced by a local translation $x'^\mu=x^\mu+a^\mu(x)$ with $a^\mu(x)=\omega^\mu_{\alpha\beta}(x)\theta^{\alpha\beta}(x)$, hence we set $f^\mu_{\alpha\beta}=\omega^\mu_{\alpha\beta}$. The condition~\eqref{Ncond} is now satisfied with $N^\mu_{\alpha\beta}=0$ provided the energy--momentum tensor $T^{\mu\nu}$ is symmetric. Eq.~\eqref{constraint} in turn implies the well-known relation for the angular-momentum tensor
\begin{equation}
M^{\mu\alpha\beta}\eqt x^\alpha T^{\mu\beta}-x^\beta T^{\mu\alpha}.
\label{Mmuab}
\end{equation}
Using the ambiguity in the definition of the Noether currents, one can ensure that this identity holds also for theories containing fields with spin. Analogous identities can be derived from relations among symmetry transformations for the Noether currents of dilations and conformal transformations~\cite{Higashijima:1994zg}.

As the last example, consider a nonrelativistic theory of a complex Schr\"odinger field $\psi(x)$. Apart from spatial translations and rotations, it is assumed to be invariant under U(1) phase transformations $\psi'(x)=e^{\imag\theta}\psi(x)$, and Galilei boosts,
\begin{equation}
\psi'(\vek x+\vek vt,t)=e^{\imag m(\vek v\cdot\vek x+\frac12\vek v^2t)}\psi(\vek x,t).
\label{galileiboost}
\end{equation}
This corresponds to Eq.~\eqref{symtransfo} with $\zeta_i=-\imag mx_i\psi$ and $\xi^\mu_i=\eta^\mu_it$; the index $i$ labels different boosts with velocities $v^i$. Observe now that a local boost can be reproduced by a combination of a local translation with $a^\mu(x)=\eta^\mu_iv^i(x)t$ and a local phase transformation with $\theta(x)=-mv^i(x)x_i$. This corresponds to $f^\mu_i(x)=\eta^\mu_it$ and $\tilde f_i(x)=-mx_i$. The consistency condition~\eqref{Ncond} then implies
\begin{equation}
T^{0i}-mj^i=\de_\mu N^{\mu i},
\label{T0i}
\end{equation}
where $T^{\mu i}$ and $j^\mu$ are the momentum current and the particle number current, respectively. The Galilei boost current reads, in accord with Eq.~\eqref{constraint},
\begin{equation}
B^{\mu i}\eqt tT^{\mu i}-mx^ij^\mu-N^{\mu i}.
\label{Bmui}
\end{equation}
These identities were established using only fairly weak assumptions. In the next subsection, we will see under what conditions the term $N^{\mu i}$ actually vanishes.

The relations such as Eq.~\eqref{Mmuab} or \eqref{Bmui} explain the redundancy in NG bosons from the quantum-field-theoretic point of view. The vertex functions of related currents share the pole structure, hence the Ward--Takahashi identities for several broken symmetries can be saturated by a single NG mode~\cite{Higashijima:1994zg}.

%%%%%%%%%%%%%%%%%%%%%%%%%%%%%%%%%%%%%%%%%%%%%%%%%%%%%%%%%%

\subsection{Noether constraints from gauged actions}

Global symmetry ensures that Noether currents satisfy the relation~\eqref{constraint} but the term $N^\mu_\alpha$ is a priori undetermined. This becomes aggravating when a microscopic theory is compared to the EFT for its low-energy degrees of freedom. Both must share the same global symmetry, but since they can be based on very different degrees of freedom (as quantum chromodynamics and its low-energy EFT, the chiral perturbation theory), there is no way to relate their $N^\mu_\alpha$s based on global symmetry alone.

The way around this problem is to couple both theories to a set of background gauge fields $A_\mu$; their generating functionals in terms of these fields then have to coincide. Technically, we replace the action $S[\phi]$ with a new action $\tilde S[\phi,A]$ such that $\tilde S[\phi,0]=S[\phi]$, which is invariant under a simultaneous \emph{local} transformation of $\phi$ and $A_\mu$. If we manage to derive the Noether currents based on the transformation rules for the gauge fields alone, we can ensure that they satisfy the same identities in the EFT as in the microscopic theory.

We will not develop the general theory whose details closely parallel the derivation of Eq.~\eqref{constraint}. Instead, we will outline the main idea and give an illustrative example. The variation of the action can be split into contributions of the matter fields and the gauge fields, $\delta\tilde S=\delta_\phi\tilde S+\delta_A\tilde S$. Using the assumed gauge invariance and setting $A=0$ so that $\delta_\phi\tilde S$ becomes equal to Eq.~\eqref{deltaS}, the Noether current can be extracted from
\begin{equation}
\int\dd x\,J^\mu(x)\de_\mu\eps(x)=-\delta_A\tilde S\Bigr\rvert_{A=0},
\end{equation}
that is, as minus the coefficient of $\de_\mu\eps$ in $\delta_A\tilde S$. The fact that two different global symmetries have the same local form means that they can be gauged simultaneously by adding a single gauge field. For instance, local translations and local rotations are both included in the group of local diffeomorphisms; the spacetime metric plays the role of the background gauge field here. This naturally leads to relations of the type~\eqref{constraint}, for the two Noether currents are obtained by functional differentiation of the action with respect to the same gauge field.

As a concrete example, note that under certain conditions, global spatial translations in a nonrelativistic theory can be promoted to nonrelativistic general coordinate invariance~\cite{Son:2005rv}. This necessarily also includes the Galilei boosts~\eqref{galileiboost} and the local U(1) phase transformations of the field $\psi(x)$. The extended action $\tilde S[\psi,g,A]$ depends on the matter field $\psi(x)$, the spatial metric $g_{ij}(x)$ and the U(1) gauge field $A_\mu(x)$. A combined local U(1) rotation with parameter $\theta(x)$ and local spatial translation with parameter $\xi^i(x)$ takes the form~\cite{Son:2005rv}
\begin{equation}
\begin{split}
\delta g_{ij}&=-\xi^k\de_kg_{ij}-g_{ik}\de_j\xi^k-g_{kj}\de_i\xi^k,\\
\delta A_0&=\de_0\theta-\xi^j\de_jA_0-A_j\de_0\xi^j,\\
\delta A_i&=\de_i\theta-\xi^j\de_jA_i-A_j\de_i\xi^j+mg_{ij}\de_0\xi^j.
\end{split}
\label{coordtrandsfo}
\end{equation}
The local phase transformation is obtained by setting $\xi^i=0$, i.e.~$\delta g_{ij}=0$ and $\delta A_\mu=\de_\mu\theta$. The variation of the action becomes
\begin{equation}
\delta\tilde S=\delta_\psi\tilde S+\int\dd x\,\VAR{\tilde S}{A_\mu}\de_\mu\theta.
\end{equation}
From here we immediately obtain the expression for the particle number current, $j^\mu\eqt-\delta\tilde S/\delta A_\mu\bigr\rvert_{A=0}$. (We denote by $A=0$ the limit of both $g_{ij}\to-\delta_{ij}$ and $A_\mu\to0$.) Analogously, we obtain the variation of the action under local translations and thence the momentum current, defined by $\delta S=\int\dd x\,T^\mu_{\phantom\mu i}\de_\mu\xi^i$,
\begin{equation}
T^{0i}\eqt-m\VAR{\tilde S}{A_i}\biggr\rvert_{A=0},\qquad
T^{ij}\eqt2\VAR{\tilde S}{g_{ij}}\biggr\rvert_{A=0}.
\end{equation}
Finally, by setting $\theta(x)=-mx_iv^i(x)$ and $\xi^i(x)=v^i(x)t$, we reproduce local Galilei boosts with parameters $v^i(x)$, which gives the boost current, defined by $\delta S=\int\dd x\,B^\mu_{\phantom\mu i}\de_\mu v^i$,
\begin{equation}
\begin{split}
B^{0j}&\eqt mx^j\VAR{\tilde S}{A_0}\biggr\rvert_{A=0}-mt\VAR{\tilde S}{A_j}\biggr\rvert_{A=0},\\
B^{ij}&\eqt mx^j\VAR{\tilde S}{A_i}\biggr\rvert_{A=0}+2t\VAR{\tilde S}{g_{ij}}\biggr\rvert_{A=0}.
\end{split}
\end{equation}
Putting together all the intermediate expressions gives the result
\begin{equation}
T^{0i}\eqt mj^i,\qquad
B^{\mu i}\eqt tT^{\mu i}-mx^ij^\mu.
\end{equation}
The reason why we obtained stronger relations than in Eqs.~\eqref{T0i} and \eqref{Bmui} is the \emph{assumption} that the theory can be made generally coordinate invariant by introducing background gauge fields $g_{ij}$ and $A_\mu$. This is a nontrivial assumption which imposes stronger constraints on the action than mere global U(1), translational and Galilei invariance~\cite{Son:2005rv}.

%%%%%%%%%%%%%%%%%%%%%%%%%%%%%%%%%%%%%%%%%%%%%%%%%%%%%%%%%%

\section{Conclusions}
\label{sec:conclusions}

In this paper, we gave a new interpretation of the inverse Higgs constraints, commonly employed in constructions of effective Lagrangians for spontaneously broken spacetime symmetries. We showed that this procedure is less arbitrary than it seems: the number of true massless NG bosons depends solely on the symmetry-breaking pattern; all nonlinear realizations of the broken symmetry can be obtained by augmenting the EFT for these NG modes with matter fields. The spontaneously broken ``redundant'' symmetries can in some systems manifest themselves by the presence of massive modes in the spectrum.

Together with recent progress in understanding of broken symmetries in the nonrelativistic context~\cite{Watanabe:2011ec,Watanabe:2012hr,Hidaka:2012ym,Watanabe:2013iia}, this provides a fairly complete picture of NG bosons in an arbitrary quantum many-body system, except for the cases where translational invariance is spontaneously broken. To fill this gap represents one of the main goals of our future work. 

%%%%%%%%%%%%%%%%%%%%%%%%%%%%%%%%%%%%%%%%%%%%%%%%%%%%%%%%%%

\section*{Acknowledgments}
We are indebted to Hitoshi Murayama for numerous discussions of the topic. The work of T.B.~was funded by the Academy of Finland, grant No.~273545. H.W.~appreciates financial support from the Honjo International Scholarship Foundation. We further acknowledge the hospitality of the Kavli IPMU, and of IFT UAM-CSIC through the Centro de Excelencia Severo Ochoa Program under grant SEV-2012-0249, where parts of the work were carried out.

%%%%%%%%%%%%%%%%%%%%%%%%%%%%%%%%%%%%%%%%%%%%%%%%%%%%%%%%%%

\paragraph*{Note added} While this paper was being finished, Ref.~\cite{Endlich:2013vfa} appeared which makes the same key observation that some of the NG fields of broken spacetime symmetries can acquire a mass term. Our field redefinition based on Eq.~\eqref{IHC} makes the nature of these modes more transparent as it demonstrates that they have no attributes of NG bosons. On the other hand, the authors of Ref.~\cite{Endlich:2013vfa} observe that in some cases, the deficit in the number of true NG bosons cannot be explained by redundancy in fluctuations of the order parameter as in Sec.~\ref{subsec:microscopic}.

%%%%%%%%%%%%%%%%%%%%%%%%%%%%%%%%%%%%%%%%%%%%%%%%%%%%%%%%%%

\bibliographystyle{model1-num-names}
\bibliography{references}

\end{document}